\documentclass[twocolumn]{aastex62}

\usepackage{graphicx}
\usepackage[utf8]{inputenc}
\usepackage{amsmath,amssymb}
\usepackage{epstopdf}
\usepackage{hyperref}

\newcommand{\be}{\begin{equation}}
\newcommand{\ee}{\end{equation}}
\newcommand{\bal}{\begin{align}}
\newcommand{\eal}{\end{align}}
\newcommand{\bear}{\begin{eqnarray}}
\newcommand{\eear}{\end{eqnarray}}
\newcommand{\nn}{\nonumber}

\newcommand{\Adiab}
{Basu:04,Van_Doorsselaere:11,Poedts:11,Fan:13,Zavershinskii:20}
\newcommand{\Polytr}{Roussev:03,Petrie:07,Cohen:07,Poedts:11,Wang:15,Zhang:16,Prasad:18,Takahashi:18}

\begin{document}

\title{On the Influence of the Ionization-Recombination Processes on Hydrogen Plasma Polytropic Index}

\author{Todor~M.~Mishonov}
\affiliation{Georgi Nadjakov Institute of Solid State Physics, Bulgarian Academy of Sciences,\\
72 Tzarigradsko Chaussee Blvd., BG-1784 Sofia, Bulgaria}

\author{Iglika~M.~Dimitrova}
\affiliation{Faculty of Chemical Technologies,
University of Chemical Technology and Metallurgy,\\
8 Kliment Ohridski Blvd., BG-1756 Sofia, Bulgaria}
\affiliation{Georgi Nadjakov Institute of Solid State Physics, Bulgarian Academy of Sciences,\\
72 Tzarigradsko Chaussee Blvd., BG-1784 Sofia, Bulgaria}

\author{Albert~M.~Varonov}
\affiliation{Georgi Nadjakov Institute of Solid State Physics, Bulgarian Academy of Sciences,\\
72 Tzarigradsko Chaussee Blvd., BG-1784 Sofia, Bulgaria}

\date{8 December 2020, 18:48}

\begin{abstract}
The polytropic (adiabatic) index for pure hydrogen plasma is analytically calculated as function of reciprocal temperature and degree of ionization.
Additionally, the polytropic index is graphically represented as a function of temperature and density.
It is concluded that the partially ionized hydrogen plasma cannot be exactly polytropic.
The calculated deviations from the mono-atomic value 5/3 are measurable.
The analytical result for pure hydrogen plasma
is a test example how this approach can be extended for arbitrary
gas cocktail. \\
\end{abstract}

\section{Introduction}
The polytropes find many many applications in astrophysics 
and related fields~\citep{Horedt:2004}
and there are a lot of hints~\citep{Totten:95,Kartalev:06} for deviation
of polytropic (or adiabatic) index $\gamma_\mathrm{eff}$ from the mono-atomic value
$\gamma_a=5/3$.
However, the 
the first measurement of the adiabatic index 
in the solar corona using time-dependent spectroscopy 
of \textit{HINODE/EIS} observations
by~\citet{Van_Doorsselaere:11}
triggered systematic study of this deviation and
put in the agenda of physics of plasmas the problem of theoretical understanding.
Similar results were obtained in \citet{Poedts:11}
and in the recent papers of \citet{Prasad:18,Zavershinskii:20}.
The following study was inspired by the paper \citep{Van_Doorsselaere:11}
where the effective adiabatic index in the solar corona is measured
for the first time by time-dependent spectroscopy of \textit{HINODE}/\textit{EIS}
observations.

Let us recall \citep{Goosens:2003} some basic definitions
\be
\label{gamma_C}
\gamma_\mathrm{eff}\equiv\frac{\mathcal{C}_p}{\mathcal{C}_v},
\quad
\mathcal{C}_p\equiv\left(\partial w/\partial T\right)_p,\quad
\mathcal{C}_v\equiv\left(\partial \varepsilon/\partial T\right)_\rho
\ee
which describes relations between 
small fluctuations of the mass density $\rho^\prime$, 
pressure $p^\prime$ and temperature $T^\prime$
\be
\frac{\rho^\prime}{\rho}=\frac1{\gamma_\mathrm{eff}}\frac{p^\prime}{p}
=\frac1{\gamma_\mathrm{eff}-1}\frac{T^\prime}{T},
\ee
where $w$ and $\varepsilon$ are the enthalpy and free energy per unit mass and $\mathcal{C}_p$ and $\mathcal{C}_v$
are the heat capacities per unit mass at constant pressure $p$
or volume and mass density $\rho.$
Authors emphasize this first measurement of $\gamma_\mathrm{eff}$
and the clear deviation from the mono-atomic value $\gamma_a=5/3$
has important implications for the solar coronal physics and its modeling~
\citep{Parker:62,Roussev:03,Cohen:07,Petrie:07,Fan:13,Usmanov:16}. 
This clear deviation gives a hint that ionization-recombination processes of
minority elements as helium, carbon, oxygen and even iron can slightly influence the thermodynamic
of the coronal plasma and such a hint has already been found \citep{Basu:04},
where it was found that the adiabatic index changes near the second helium ionization. 
More hints can be found in the measurements of the adiabatic index in solar flaring loops \citep{Wang:15}, whose value is close to 5/3 and investigations of space and laboratory plasmas suggest that although the solar wind electrons have a polytropic index of less than 5/3, their actual transport might be adiabatic
\citep{Zhang:16}.
At Mega-Kelvin temperatures the solar corona hydrogen is completely ionized.
Even in the low-frequency static approximation taking into account the Saha equation requires significant amount of data and numerical calculation. 
In order to check whether such thermodynamic effects deserve to be studied in detail,
in the present comment we represent the textbook like  behavior of pure hydrogen plasma
where the same effect of deviation of adiabatic index from atomic value can be observed 
at significantly smaller temperatures, say 30~kK which correspond to the transition region.
Even from the beginning the theory should have qualitatively agreement with the experiment.

\textcolor{black}{
In order to avoid terminological misunderstandings we will
recall some basic thermodynamic relations. 
Often in hydrodynamics is used notion of liquid particle
which means small marked volume $V$ of the fluid 
with local temperature $T$ and pressure $P\equiv p$ which contains however big enough number of particles~\citep[Sec.~1]{LL6}. 
Following~\citep[Sec.~16]{LL5} we write
\begin{align}
\left(\dfrac{\partial V}{\partial P}\right)_S
&=\dfrac{\partial (V,S)}{\partial (P,S)}
=\dfrac{\dfrac{\partial (V,S)}{\partial (V,T)}}
         {\dfrac{\partial (P,S)}{\partial (P,T)}}
   \dfrac{\partial (V,T)}{\partial (P,T)}  \\  
&=\dfrac{T\left(\dfrac{\partial S}{\partial T}\right)_{\!V}}
           {T\left(\dfrac{\partial S}{\partial T}\right)_{\!P}}  
     \left(\dfrac{\partial V}{\partial P}\right)_{\!T} 
    =\dfrac{C_v}{C_p} \left(\dfrac{\partial V}{\partial P}\right)_{\!T},  
\nn      
\end{align}
where $S$ is the entropy and $C_p$ and $C_v$
are the heat capacities of the liquid particle for constant pressure and volume.
Substituting then volume via mass $M$ of the liquid particle as 
$V=M/\rho$ we obtain the relation between
ratio of heat capacities and compressibilities
\begin{align}
\gamma\equiv\dfrac{C_p}{C_v}
=\dfrac{\left(\dfrac{\partial V}{\partial P}\right)_{\!T}}
         {\left(\dfrac{\partial V}{\partial P}\right)_{\!S}}
=\dfrac{\left(\dfrac{\partial \rho}{\partial P}\right)_{\!T}}
         {\left(\dfrac{\partial \rho}{\partial P}\right)_{\!S}}
=\dfrac{\left(\dfrac{\partial p}{\partial\rho}\right)_{\!S}}
         {\left(\dfrac{\partial p}{\partial\rho}\right)_{\!T}}
=\frac{v_S^2}{v_T^2}.
\end{align}
We emphasize that this is only ratio between derivatives
and it is not supposed that plasma have polytropic 
adiabatic equation $PV^\gamma=\mathrm{const}$
which is property of ideal gas with constant heat 
capacity, see~\citet[Sec.~43]{LL5}.
Often is introduced notation
\begin{align}
\left(\frac{\partial p}{\partial\rho}\right)
_{\!\!\mathrm{equilibrium}}
\equiv 
\left(\dfrac{\partial p}{\partial\rho}\right)_{\!\!S}=v_S^2
\end{align}
emphasizing that for slow hydrodynamic and MHD processes 
plasma follows in every moment Saha equilibrium conditions
and this slow process is reversible with negligible entropy production.
Another often used notion is the fast adiabatic comprehensibility
at constant ionization degree 
\be
v_\infty^2
\equiv \left(\frac{\partial p}{\partial\rho}\right)_{\!\!\alpha}.
\ee
As we will see in the next section the heat capacity of partially 
ionized plasma is temperature and density dependent and can be much bigger than one and definitely chromospheric plasma is not polytropic.
In some articles index $\gamma$ is called 
adiabatic~\citep{\Adiab} in some polytropic~\citep{\Polytr}
but they are synonyms.
}

\section{Calculation for pure hydrogen plasma}

The purpose of the present paper is to represent analytical result for the effective 
adiabatic index $\gamma_\mathrm{eff}$ for hydrogen plasma which consists of
electrons, protons and hydrogen atoms with volume densities
$n_e$, $n_p$ and $n_0$ respectively.

The correlation energy is negligible for atmospheric plasma and with acceptable approximation
the pressure and mass density are described by the total density of the particles of an ideal gas
\begin{align}&
p=n_\mathrm{tot}T,\quad n_\mathrm{tot}=n_e+n_p+n_0,\\&
\rho=Mn_\rho,\quad n_\rho=n_0+n_p,\quad \alpha\equiv n_p/n_\rho,
\end{align}
where $M$ is the proton mass, and $\alpha$ is the degree of ionization.

\begin{figure}[h]
\includegraphics[scale=0.5]{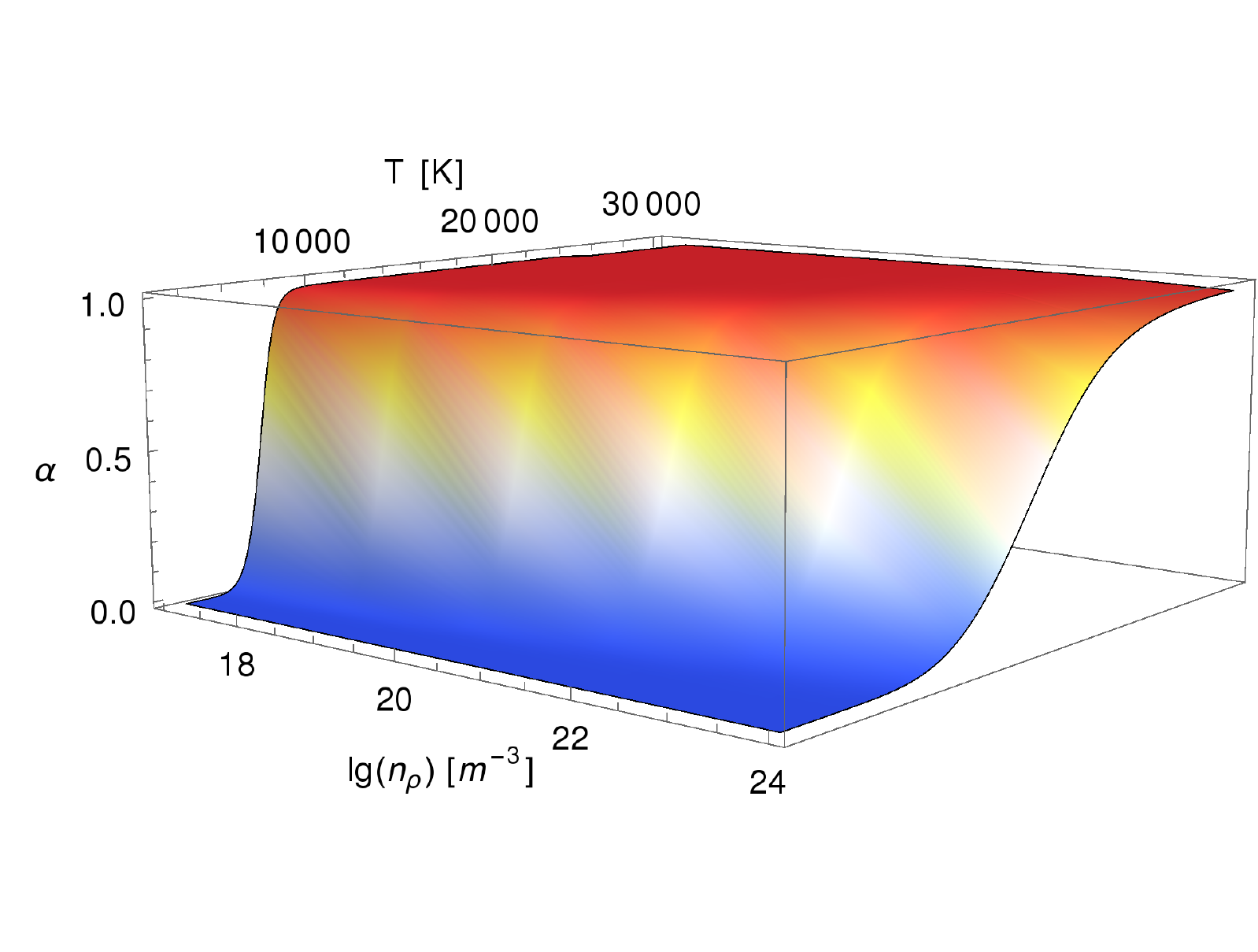}
\caption{Degree of ionization $\alpha$ in vertical direction as a function
temperature and density in logarithmic scale, i.e. as function of $T$ and $\lg n_\rho.$}
\label{Fig:alpha}
\end{figure}

The internal energy per unit mass $\varepsilon$ and the enthalpy per unit mass $w$
are given by 
\begin{align}&
\varepsilon=\left(c_{v,a} Tn_\mathrm{tot}+I n_e\right)/\rho,\quad w=\varepsilon+p/\rho,
\quad n_\rho=\rho/M, \nonumber
\\&\nonumber
n_e=n_p=\alpha n_\rho, \quad n_0=(1-\alpha)n_\rho,\quad
n_\mathrm{tot}=(1+\alpha)n_\rho,\\&\nonumber
c_{v,a}\equiv \frac32, \quad c_ {p,a}\equiv c_{v,a}+1=\frac52,\quad 
\gamma_a\equiv \frac{c_{p,a}}{c_{v,a}}=\frac53.
\end{align}
Here $c_{p,a}$ and $c_{v,a}$ are just mathematical constants
taken from the theory of mono-atomic ideal gasses.
Simultaneously the degree of ionization is given by the Saha \citep{Saha:21} equation
\begin{align}&\nn
\alpha\equiv \frac{n_p}{n_\rho}=\frac1{\sqrt{1+p/p_\mathrm{_S}}},\quad
\frac{p}{p_\mathrm{_S}}=\frac1{\alpha^2}-1,\quad 
\frac{n_\rho}{n_\mathrm{_S}}=\frac{1-\alpha}{\alpha^2},
\\&
p_\mathrm{_S}\equiv n_\mathrm{_S}T,\quad
n_\mathrm{_S}\equiv n_q\mathrm{e}^{-\iota},\quad\iota\equiv\frac{I}{T},\quad
n_q=\left(\frac{mT}{2\pi\hbar^2}\right)^{\!\! 3/2},\nn
\end{align}
where $I$ is the hydrogen ionization potential and $m$ is the electron mass.
The dependence $\alpha(T/k_\mathrm{B},n_\rho)$ is given 
in Fig.~\ref{Fig:alpha}.
The degree of ionization $\alpha$ depends on the density and pressure and that is why the 
internal energy and enthalpy obtain pressure and mass density dependence
\begin{align}
& \varepsilon=\frac1{M} \left[c_{v,a}(1+\alpha)T+I\alpha\right],\\
& w= \frac1{M} \left[c_{p,a}(1+\alpha)T+I\alpha\right].
\end{align}
We have to emphasize that magnetic pressure and magnetic field in general
exactly zero influence on the thermodynamic properties of classical plasma.
This result is known as Bohr–Van~Leeuwen theorem,
see the well-known monograph by~\citet[Chap.~1]{Mattis:65} and
the cited therein monograph by~\citet[Chap.~4, Sec.~24, p.~94]{VanVleck}
and PhD theses by~\cite{Bohr} and~\cite{VanLeeuwen}.
Taking the differential from the expression for $\alpha$
the calculation gives
\begin{align}&
T\left(\frac{\partial\alpha}{\partial T}\right)_{\!\!p}
=\frac{(1-\alpha^2)\alpha}{2}\,(c_{p,a}+\iota),\\&
T\left(\frac{\partial\alpha}{\partial T}\right)_{\!\!\rho}
=\frac{(1-\alpha)\alpha}{2-\alpha}\,(c_{v,a}+\iota).
\end{align}
Further differentiation of the thermodynamic potentials with respect to the temperature according to
Eq.~(\ref{gamma_C}) gives 
\begin{align}&
\tilde{c}_p\equiv
\frac{\rho\,\mathcal{C}_p}{n_\mathrm{tot}}
=c_{p,a}+(c_{p,a}+\iota)^2\varphi,\\&
\tilde{c}_v\equiv\frac{\rho\,\mathcal{C}_v}{n_\mathrm{tot}}
=c_{v,a}+(c_{v,a}+\iota)^2\varphi/(1+\varphi),\\&
\tilde\gamma=\dfrac{\mathcal{C}_p}{\mathcal{C}_v}
=\frac{c_{p,a}+(c_{p,a}+\iota)^2 \varphi}
{c_{v,a}+(c_{v,a}+\iota)^2 \varphi/(1+\varphi)},
\label{gamma}\\&
\varphi\equiv\frac12(1-\alpha)\alpha, \quad 
\frac{\rho}{n_\mathrm{tot}}=\left<M\right>\equiv\frac{M}{1+\alpha},
\end{align}
where $\left<M\right>$ is the averaged mass of the cocktail,
and $\tilde{c}_v(\iota,\alpha)$ and $\tilde{c}_p(\iota,\alpha)$ 
are temperature and ionization dependent 
heat capacities per particle; the temperature is in energy units.

\begin{figure}[h]
\includegraphics[scale=0.5]{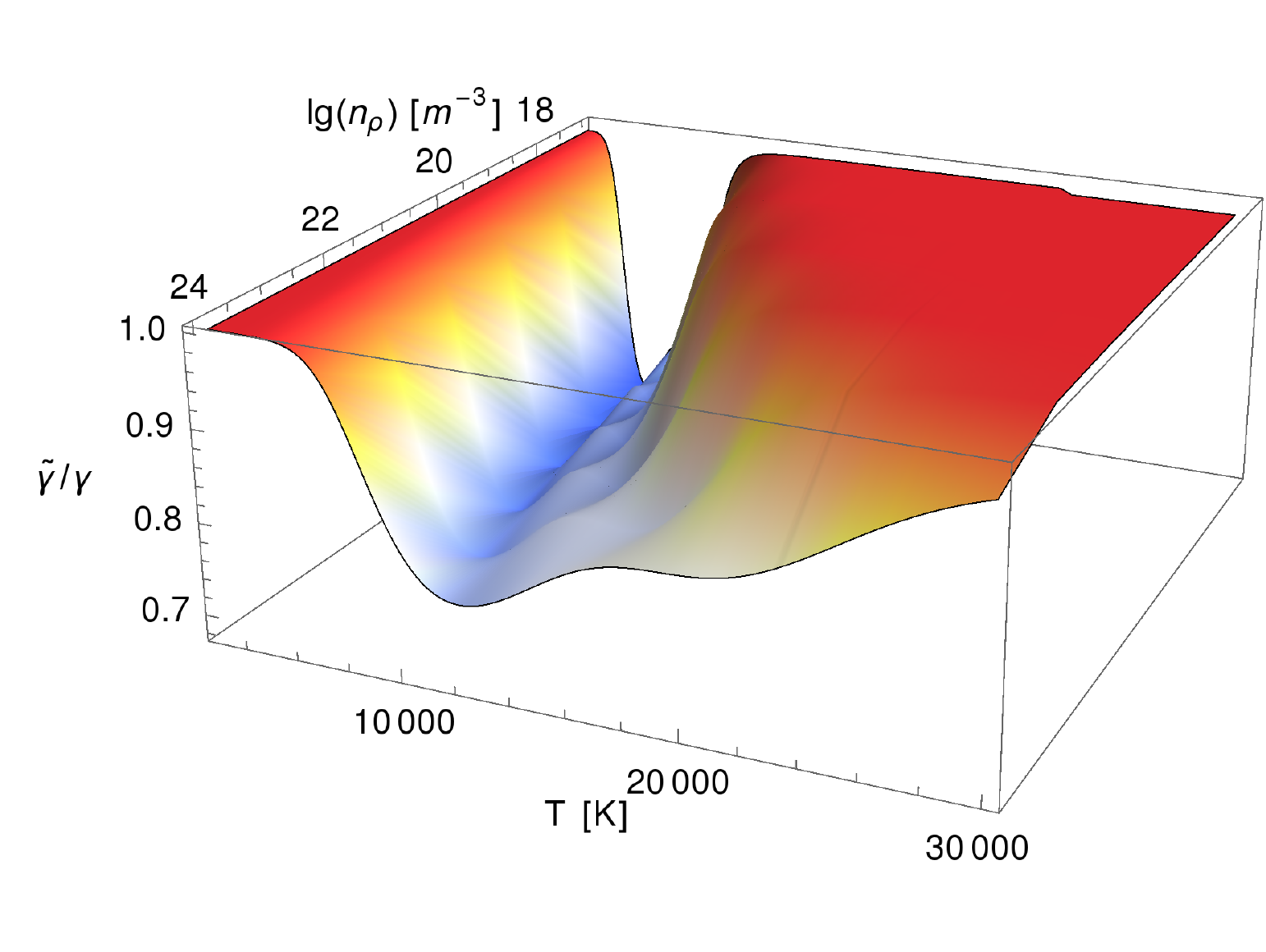}
\caption{Relative adiabatic index $\tilde\gamma/\gamma_a$ as a function of
temperature and density in logarithmic scale; 
$\tilde\gamma \equiv \gamma_\mathrm{eff}$.
The temperatures correspond from around the solar photosphere to the transition region.
We expect that partial ionization of heavy elements 
will give similar behavior in the solar corona for the higher temperatures.}
\label{Fig:gamma}
\end{figure}

\begin{figure}[h]
\includegraphics[scale=0.33]{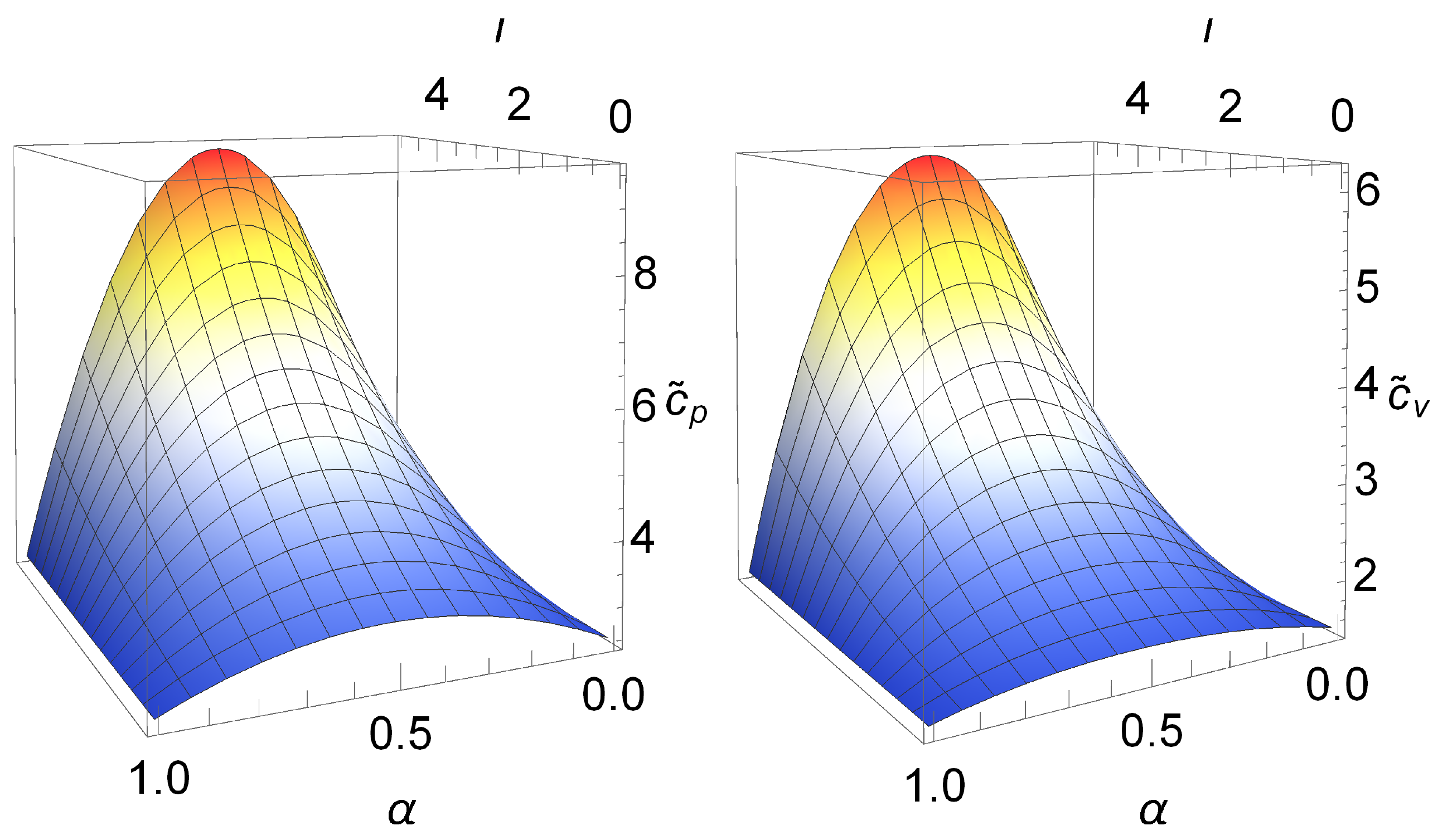}
\caption{Heat capacity per particle at constant pressure 
$\tilde{c}_p(\iota,\alpha)$ (left)
and heat capacity at constant volume
$\tilde{c}_v(\iota,\alpha)$ (right)
per particle of partially ionized pure hydrogen plasma.
One can see significant increase of both heat capacities at small temperatures $T$ related to energy of ionization $I$ of the plasma and both heat capacities have almost identical behavior with the only clearly visible difference being the scales of the vertical direction. 
}
\label{Fig:c_p_c_v_iota_alpha}
\end{figure}

\begin{figure}[h]
\includegraphics[scale=0.45]{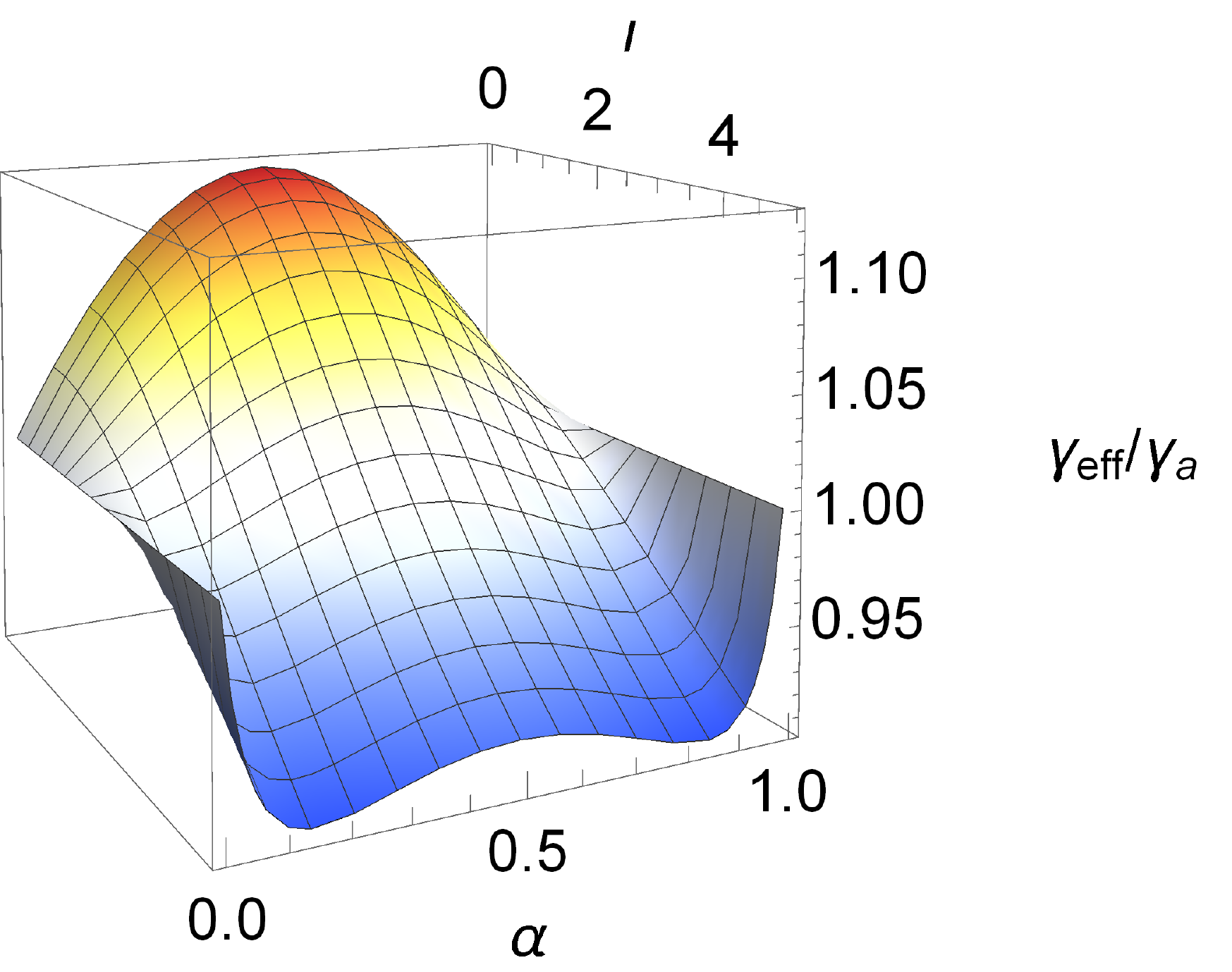}
\caption{
Analytical result for relative polytropic index 
$\gamma_\mathrm{eff}/\gamma_a$
as a function of degree of ionization $\alpha$ and reciprocal temperature $\iota\equiv I/T$
for pure hydrogen plasma.
}
\label{Fig:gamma_iota_alpha}
\end{figure}

One can see in Fig.~\ref{Fig:gamma} that the relative adiabatic index $\tilde\gamma/\gamma_a$
can differ significantly from 1 even when the temperature is high enough 
and the degree of ionization is almost 1.
The dependency $\tilde\gamma/\gamma_a$ in Fig.~\ref{Fig:gamma}
is shown only up to 30~kK temperature,
which roughly corresponds to the beginning of the solar transition region \citep{Eddy:79,Avrett:08}
in order the deviation from 1 to be seen in detail.
For higher temperatures its value is clearly 1, which is well-known and of course anticipated since we have included only pure hydrogen in our treatment.
Our analytical results for the heat capacities
$\tilde c_p$, $\tilde c_v$ and 
their ratio $\gamma_\mathrm{eff}=\tilde c_p/\tilde c_v$
are depicted in 
Figs.~\ref{Fig:c_p_c_v_iota_alpha} and \ref{Fig:gamma_iota_alpha}.
\textcolor{black}
{
This correction will not change qualitatively the
uncountable MHD simulations but let be quantitatively correct.
MHD is science not a model and the nature of the effective polytropic index
was discussed in the excellent monograph by~\citet{Goosens:2003}.
In great detail hydrodynamics and MHD of fluid with chemical reactions was discussed also in the 
monographs~\citet[Chap.~12]{deGroot:74}, \citet[Sec.~4]{Rudenko:77} and references therein.
}
Both heat capacities have almost identical behavior, the only visible difference being the vertical scales.
The symmetry of the heat capacities and the relative polytropic index about the maximal value $\alpha=0.5$ is governed by 
$\varphi$.
Despite the large values of the heat capacities, their quite similar behavior limits the values of the relative polytropic index to within around 10\% of $\gamma_a$.

\section{Conclusions}
\textcolor{black}
{
Let us summarize the novelty of our results.
We have derived for the first time explicit expressions
for the adiabatic index and heat capacities of pure hydrogen plasma.
Our results are directly applicable for the solar chromospere
where ionization-recombination processes of heavy elements have
negligible contribution.
Our approach is also applicable for pure 
argon~\citep{Takahashi:18}, the solar corona and arbitrary plasma cocktail.
It is necessary to solve the corresponding Saha equations 
for all ions in the fluid and to substitute the obtained concentrations in the expressions for the enthalpy and free energy.
Calculation of the heat capacities and polytropic index $\gamma$ 
is then a doable task using the latest atomic database CHIANTI~\citep{Dere:19}.
Analogously for tokamak plasma evaporation of small amount
of the material from the panels significantly decreases plasma temperature due to ionization of heavy elements.
And the polytropic index again will be different from 
the single-atomic value $\gamma_a=5/3$ from the computer simulations.
}

The solar corona and stellar atmospheres in general 
contain heavy elements and even ionization of helium
can create significant changes of the polytropic index \citep{Basu:04}.
It is a routine task for every plasma cocktail to include the
Saha ionization equation in its thermodynamics.
For pure argon used in the laboratory
experiments \citep{Takahashi:18} the task is even simplified.

In conclusion, we consider that the experimental data processing of the astrophysical observations
has to start with the equilibrium thermodynamics of 
realistic chemical compound for which is possible
to make state of the art theoretical evaluation of $\gamma_\textrm{eff}$.
Our analytical result for pure hydrogen plasma Eq.~(\ref{gamma})
is just the illustration of the first step.
And this first step is a necessary ingredient for the explanation of the physical processes in the solar chromosphere, for instance what causes the hydrogen ionization there.

The next problem of the physics of solar corona is to recalculate the dispersion relations 
of magneto-hydrodynamic waves taking into account the influence of ionization-recombination processes on the kinetic coefficients.
Wave propagation and kinetic effects related to frequency dependent misbalance
requires even more sophisticated treatment.
For example, even the second viscosity of the hydrogen plasma and its dispersion is still an open problem 
in astrophysics.

\textit{Acknowledgments.} 
The authors thanks to Valery Nakariakov and Kris Murawski
for the interest to the paper, correspondence and valuable remarks.

\textit{Apropos:}
The experimental set-up presented in Fig.~1 of the commented article \citet{Takahashi:18} remains a propulsion engine of a magneto-plasma rocket.
We use the opportunity to mention 
a new idea that not only helicon waves 
but antennas exciting Alfv\'en waves (AW) 
can be even the better solution
for heating of hot dense plasma by viscosity friction. 
The area of of AW damping will be similar to the combustion chamber of chemical jet engines. 
And creation of propulsion will be analogous to the launching of solar wind by absorption of AW as Hannes Alfv\'en suggested many years ago \citep{Alfven:42,Alfven:47}.


\bibliography{gamma}
\end{document}